\documentclass[10pt,conference]{IEEEtran}
\IEEEoverridecommandlockouts

\usepackage[bookmarks=false]{hyperref}
\usepackage{lipsum}
\usepackage{hyperref}
\usepackage{url}
\usepackage{amssymb}
\usepackage{rotating}
\usepackage{multirow}
\usepackage{array}
\usepackage{amssymb}
\usepackage{amsfonts}
\usepackage{graphicx}
\usepackage{epstopdf}
\usepackage{amsmath}
\usepackage{booktabs}
\usepackage{epigraph}
\usepackage{listings}
\usepackage{color}
\usepackage{wasysym}
\usepackage{float}
\usepackage{amsmath}
\usepackage{pifont}
\usepackage[table,xcdraw]{xcolor}

\usepackage{booktabs}
\usepackage{listings}
\usepackage{color}
\usepackage{xspace}
\usepackage{svg}
\usepackage{framed}
\usepackage{flushend}

\definecolor{shadecolor}{rgb}{0.75, 0.75, 0.75}

\newcommand*{\coveralls}{\textsc{Coveralls}\xspace}
\newcommand*{\travist}{\textsc{TravisTorrent}\xspace}
\newcommand*{\travisci}{\textsc{TravisCI}\xspace}

\definecolor{dkgreen}{rgb}{0,0.6,0}
\definecolor{gray}{rgb}{0.5,0.5,0.5}
\definecolor{mauve}{rgb}{0.58,0,0.82}

\makeatletter
\newcommand\footnoteref[1]{\protected@xdef\@thefnmark{\ref{#1}}\@footnotemark}
\makeatother

\newcommand{\MyBox}[1]{\vspace{3mm}\noindent\framebox[\columnwidth][c]{\parbox[b]{0.95\columnwidth}{ #1 }}}

\begin{document}

\title{Continuous Integration Theater}

% \author{Anonymous}

\author{\IEEEauthorblockN{Wagner Felidr\'e, Leonardo Furtado}
\IEEEauthorblockA{UFPA \\ Bel\'em, Brazil\\
\{wagnerfelidre,srleonardofurtado\}@gmail.com}
\and
\IEEEauthorblockN{Daniel A. da Costa}
\IEEEauthorblockA{University of Otago \\ Dunedin, New Zeland\\
danielcalencar@otago.ac.nz}
\and
\IEEEauthorblockN{Bruno Cartaxo}
\IEEEauthorblockA{IFPE \\ Paulista, Brazil \\ email@brunocartaxo.com}
\and
\IEEEauthorblockN{Gustavo Pinto}
\IEEEauthorblockA{UFPA \\ Bel\'em, Brazil\\gpinto@ufpa.br}
}

\IEEEpubid{\makebox[\columnwidth]{978-1-7281-2968-6/19/\$31.00~\copyright2019 IEEE \hfill} \hspace{\columnsep}\makebox[\columnwidth]{ }}

\maketitle

\IEEEpubidadjcol

\begin{abstract}
\textit{Background}: Continuous Integration (CI) systems are now the bedrock of several software development practices. Several tools such as TravisCI, CircleCI, and Hudson, that implement CI practices, are commonly adopted by software engineers. However, the way that software engineers use these tools could lead to what we call ``Continuous Integration Theater'', a situation in which software engineers do not employ these tools effectively, leading to unhealthy CI practices.
\textit{Aims}: The goal of this paper is to make sense of how commonplace are these unhealthy continuous integration practices being employed in practice.
% to software systems less robusts, when it comes to dealing with frequent changes. \\
\textit{Method}: By inspecting 1,270 open-source projects that use TravisCI, the most used CI service, we quantitatively studied how common is to use CI (1) with infrequent commits, (2) in a software project with poor test coverage, (3) with builds that stay broken for long periods, and (4) with builds that take too long to run. 
\textit{Results}: We observed that 748 ($\sim$60\%) projects face infrequent commits, which essentially makes the merging process harder. Moreover, we were able to find code coverage information for 51 projects. The average code coverage was 78\%, although Ruby projects have a higher code coverage than Java projects (86\% and 63\%, respectively). However, some projects with very small coverage ($\sim$4\%) were found.
Still, we observed that 85\% of the studied projects have at least one broken build that take more than four days to be fixed. Interestingly, very small projects (up to 1,000 lines of code) are the ones that take the longest to fix broken builds. 
Finally, we noted that, for the majority of the studied projects, the build is executed under the 10 minutes rule of thumb.
% Finally, we observed that 1,255 of the studied projects that 2 days to fix a broken build. \\
\textit{Conclusions}: Our results are important to an increasing community of software engineers that employ CI practices on daily basis but may not be aware of bad practices that are eventually employed.

\end{abstract}

\begin{IEEEkeywords}
	Continuous Integration, Test coverage, Bad practices
\end{IEEEkeywords}

\section{Introduction}\label{sec:intro}

Continuous Integration (CI) is the practice of merging all developer working copies into a shared mainline, several times a day~\cite{Humble:2010}. Although the culture of continuously integrating changes dates from the 70s~\cite{Brooks:1978}, CI practices has gained momentum only in the last 10 years, being more widely discussed, employed, and researched. Consequently, CI is nowadays one of the pillars of the software engineering practice, not only in commercial projects, but also in open source projects~\cite{Pinto:SPE:2018, Reboucas:2017:CII}. 

The success of CI can be partially accredited to first world class tools that have considerably automated most of the required steps to inspect, integrate, and test source code change in a transparent and straightforward manner. The use of CI tools not only accelerates the software development process (since merging changes become more frequent without reducing software quality~\cite{Vasilescu:2015:QPO}), but software bugs can also be identified earlier and faster~\cite{Vasilescu:2014:CIS}.

There are several tools offering support for developers that plan to incorporate the CI practices into their software projects. Such tools include TravisCI, CircleCI, and Hudson. More interestingly, however, is the fact that some of these tools are readily available in social coding environments such as GitHub and GitLab. Essentially, this integration implies that everyone with a GitHub account can gratuitously benefit from the complex pipeline of version control systems, code review systems, and continuous integration tools, with little to no configuration effort. Therefore, it comes as no surprise to state that CI tools (e.g., TravisCI) are highly used and demanded by developers~\cite{Hilton:2016}.

% Continuous Integration (CI) nowadays is an agile practice widely employed in not only commercial projects, but also open source projects. 

However, the sole usage of CI tools does not necessarily imply that a software development team properly adhere to the CI practices. Recent works have shown that the use of automation tools may produce no benefits, unless the development team is willing to change their development culture~\cite{Luz:ESEM:2018}. As an example, according to Fowler~\cite{fowler:CI}, one of the CI practices is to promote self-testing builds. However, to have a self-testing code one needs a suite of automated tests that can check a large part of the code base for eventual bugs. Unfortunately, although CI tools re-execute test suites after every new change (i.e., to avoid the introduction of new bugs), CI tools cannot identify whether the software project contains a comprehensive test suite. Therefore, in such a case, the development team will not benefit from the test automation supposed to be promoted by CI tools. Thus, a common misconception that has been acknowledged about CI is that the sole adoption of a CI tool does not imply the proper adherence to CI practices~\cite{Luz:ESEM:2018,Zhang:FSE:2018}. 
Indeed, such kind of situation has long been one of the Achilles' heels of agile. To publicize that is adherent to agile practices aiming to gain some kind of credibility, while under the covers the basic practices are not properly followed~\cite{ELORANTA2016}.

In this paper, we investigate a set of CI bad practices. These bad practices are related to the use of CI (1) with infrequent commits on the master branch (i.e., delaying integration), (2) in a software project with poor test coverage (i.e., missing eventual bugs), (3) with builds that remain broken for long periods for time (i.e., blocking new features), and (4) with builds with considerably long durations (i.e., limiting the rapid feedback). These bad practices constitute what is known as the ``Continuous Integration Theater''\footnote{https://www.thoughtworks.com/radar/techniques/ci-theatre} in the practitioners arena. According to the grey literature:

\begin{shaded}
\centering
\noindent
``Continous Integration Theater describes the illusion of practising continuous integration while \\ not really practising it.''\footnote{https://www.gocd.org/2017/05/16/its-not-CI-its-CI-theatre.html}.
\end{shaded}

Although these bad practices are commonly discussed in the grey literature, little research has been devoted to shed some light on the existence of projects performing the CI Theater. 

To conduct this investigation, we leverage the \travist dataset, which is a comprehensive dataset of data and metadata regarding projects that use \travisci. Whenever necessary, we enriched this dataset with data from \coveralls, which is a third-party service that provides test coverage information. Through a mostly quantitative analysis over 1,270 open source projects and their 534,417 builds, we produce a list of findings regarding CI bad practices that are employed in open source projects, some of which are not always obvious. We now highlight our main findings here.

\begin{enumerate}
    \item[\textbf{RQ1)}] \textbf{Infrequent commits are frequent.} We empirically defined the value of our metrics for infrequent commits as 2.36 commits per weekday. We then found that 60\% of the studied projects have less than 2.36 commits, suffering from infrequent commits. The size of the project has no influence on the (in)frequency of commits. Large Ruby projects, however, are the most active ones and do not adhere to this rule. 
    \item[\textbf{RQ2)}] \textbf{Test coverage could mislead CI results.} We identified 51 projects that we could measure test coverage information. On average, Java projects have 63\% of test coverage, whereas Ruby projects have 86\%. At the bare minimum, we find one Java project with 4\% of test coverage, and one Ruby project with 14\% of test coverage. This finding suggests that the report of a CI service could be compromised, since some projects might not place enough care in curating their test coverage (e.g., a passing build may be hiding bugs due to the poor test coverage). 
    \item[\textbf{RQ3)}] \textbf{Long to be fixed broken builds.} We observed that 85\% of the analyzed projects have at least one build that took more than four days to be fixed. This finding is particularly unfortunate since broken builds that take several days to be fixed may introduce an additional burden (or distrust) on the development team. Interestingly, we observed that large projects (either Java or Ruby) have less instances of long to be fixed broken builds than smaller projects. These long to be fixed builds, on very small projects, are fixed, on average, in 40 days, which is strong smell of the CI theater.
    \item[\textbf{RQ4)}] \textbf{Builds are executed quickly, though.} In order to provide quick feedback, builds should be executed under 10 minutes~\cite{fowler:CI,Hilton:FSE:2017}. We found only 43 projects that do not adhere to this general rule of thumb. As an exception to this rule, we found 43 very large and complex projects, such as  the JRuby (the Ruby implementation for the Java VM) or the Facebook Presto (a distributed SQL query engine for big data), that have builds which take longer than 30 minutes. In spite of these cases, this symptom of the CI theater was hardly observed.
\end{enumerate}

% In this work we studied a set of Continuous Integration bad practices.  

%Developers can configure CI services,such as Travis, to post code coverage data to a service such as Coveralls, which makes the data easily available to developers,for instance when reviewing pull requests. \bnote{paragrafo ta sem conexao com o resto do texto}

\section{Method}\label{sec:method}

In this section we introduce our research questions (Section~\ref{sec:rqs}, and the approaches we used to gather (Section~\ref{sec:dataset}), analyze (Section~\ref{sec:analysis}) data. We also provide a package for help anyone who want to replicate this study(Section~\ref{sec:replication}).

% In this section we introduce our research questions (Section~\ref{sec:rqs}, and the approaches we used to gather (Section~\ref{sec:dataset}), analyze (Section~\ref{sec:analysis}), and to help other replicate our study (Section~\ref{sec:replication}).

\subsection{Research Questions}\label{sec:rqs}

In this work we studied the following four important research questions. 

\MyBox{\textbf{RQ1:} How common is running CI in the master branch but with infrequent commits?}

\vspace{0.2cm}
\noindent
\textbf{Rationale.} One of the main advantages of CI systems is that they decrease the pain of merging new changes. This relief comes from the practice of merging continuously. However, sometimes software engineers opt not to integrate continuously (e.g., they take too much time working on a separate branch and only after days of work they apply the changes in the master branch). Practitioners have baptized the bad practice of working in silos---either in their local branches or remote branches---as ``Continuous Isolation''\footnote{https://medium.com/continuousdelivery/continuous-integration-not-continuous-isolation-d068a756df0f}. 

%However, this relief comes from the practice of merging continuously. When software engineers opt not to integrate continuously (e.g., they take too much time working on a separate branch and only after days of work they apply the changes in the master branch). Practitioners have baptized the bad practice of working in silos either in their local branches or remote branches as ``Continuous Isolation''. 

\MyBox{\textbf{RQ2:} How common is running a build in a software project with poor test coverage?}

\vspace{0.2cm}
\noindent
\textbf{Rationale.} Test coverage measures how much of a software project is exercised during testing. If a project has a fragile test suite (and consequently a low test coverage), new changes that clearly introduce bugs are potentially not caught during build time. Therefore, CI systems offer little help in software projects that do not carefully build their testing arsenal. Although many criteria were introduced to measure code coverage~\cite{Ammann:2002:ISSRE,Gligoric:2013:ISSTA,Atanas:2005:FASE}, roughly speaking, test coverage is measured by the number of lines of code exercised by test cases divided by the total number of lines of code. %\bnote{e qual o threshold pra definir como poor test coverage?}

% Test coverage measures how much of the software project is  exercised during testing. If a project has poor test coverage, new changes that clearly introduce bugs are potentially not caught during the build time. Therefore, CI systems offer little help in software projects that do not carefully build their testing arsenal.

\MyBox{\textbf{RQ3:} How common is allowing the build to stay broken for long periods?}

\vspace{0.2cm}
\noindent
\textbf{Rationale.} Here we sought to investigate how common and how long broken builds stay broken in our dataset. A broken master is particularly undesirable because it may block features from rolling out (i.e., a faulty commit needs to be detected and rolled backed). Notable practitioners, such as Martin Fowler, have suggested that ``if the mainline build fails, it needs to be fixed right away''~\cite{fowler:CI}, making a broken build an urgent, high priority task. However, if broken builds stay red longer than this, it may suggest that projects maintainers may not be taking into account the build status and, perhaps, releasing software with bugs. Still, if developers work on a faulty master, their productivity may get hampered substantially.

\MyBox{\textbf{RQ4:} How common are long running builds?}

\vspace{0.2cm}
\noindent
\textbf{Rationale.} In this final research question, our intention is to explore how long take the builds in our dataset to process. The whole point of Continuous Integration is to provide rapid feedback. Advocates from the XP practices provide a general rule of thumb suggesting that, for most projects, 10 minutes is an expected metric. According to Fowler, ``it's worth putting in concentrated effort to make [the ten minutes rule] happen, because every minute you reduce off the build time is a minute saved for each developer every time they commit.''~\cite{fowler:CI}

\subsection{Curating the Dataset}\label{sec:dataset}

To conduct this research, we rely mostly on the dataset curated by \travist~\cite{Beller:MSR:2017}. This dataset focus on software builds created and reported in the \travisci platform, which is one of the most popular CI services nowadays. As of 2017, \travisci was reported being present in 50\% of the projects hosted on GitHub\footnote{https://github.blog/2017-11-07-github-welcomes-all-ci-tools/}. The most recent release of \travist is from November, 1st, 2017. More concretely, this dataset stores information about the builds executed, the build logs, how many tests were executed (and which ones failed), etc.
Although the last release of the datasets is from 2017, we observed that the dataset contains build information between February 2012 and March of 2016. The initial status of the dataset contains information about 1,283 open source projects, 3,702,595 build jobs, and 3,702,595 commits.
We performed three additional filtering steps in the dataset, namely:

\begin{itemize}
    \item \textbf{Removing Not a Number (NaN) records.} When analyzing the dataset, we noticed that there are some inconsistencies between the number of builds and the number of commits. Since the relationship between commit and build is one to one, we found puzzling cases in which there are more commits than builds. When analyzing the dataset, we perceived the existence of some NaN records, which our script computed as zero. We inquired \travist documentation, and it informs, in rare cases, that the \travist infrastructure does not record a push event for every build confirmation, thus generating the aforementioned data inconsistency. We then removed the rows that have NaN columns. 
    % This reduced the dataset to: 1,274 projects, 2,948,733 build jobs, and 7,081,462 commits. 
    
    % \item \textbf{Removing Not a Number (NaN) records.} When analyzing the dataset we noticed that there were some inconsistencies between number of builds and the number of commits. Since the relationship between commit and build is one to one, we found strange cases in which we had more commits than builds. When analyzing the dataset, we perceived the existance of some NaN records, which our script computed as zero. We consulted the \travist documentation, and they informed that, in rare cases, the \travist infrastructure does not record a push event for every build confirmation, thus generating the aforementioned data inconsistency. We then removed the rows that had NaN columns. This reduced the dataset to: 1,274 projects, 2,948,733 build jobs, and 7,081,462 commits. 
    
    %\gnote{Wagner, não entendi essa diferença entre numero de commits e quantidade de commits. escreva novamente em portugues} \lnote{Percebemos que a quantidade builds (Linhas do dataset)  era diferente da quantidade de registros de commits (Registro usado na coluna), isso acontece na hora de pegar os dados dessa coluna, ele procura por n\'umeros, e NaN n\~ao \'e um n\'umero. Logo pode ter 4 linhas de builds, mas apenas 3 delas est\~ao preenchidas com n\'umeros.}
    
    %This data encompasses the period from September, 5th, 2011 to November, 4th, 2015.\gnote{mas o dataset não é de 2017? como o ultimo registro é até 2015?}
    
    \item \textbf{Removing duplicated jobs.} We noted that some projects are configured to test the build against several different configurations (jobs). Since studying different jobs is not part of the scope of this research, we decided to remove duplicated jobs. 
    
    % This reduced the dataset to: 1,274 \bnote{continuou com a mesma quantidade de projetos do fitro anterior (NaN)?} projects, 536,047 build jobs, 1,295,741 commits. 
    
    %\merge{A quantidade de commits foi reduzida visto que haviam varias linhas com a mesma build ao remover essas linhas houve influencia na quantidade de commits do dataset visto que uma build poderia possuir mais de 5 linhas com o mesmo commit.}
    
    \item \textbf{Removing JavaScript projects.} After performing these two filters, we noted that only four projects were written in JavaScript, namely \texttt{zhangkaitao/es}, \texttt{dianping/cat}, \texttt{palantir/eclipse-typescript}, and \texttt{brooklyncentral/clocker}. We opted to do not consider JavaScript projects due to the small sample.
    
    %\gnote{quais sao esses projetos?}
    
\end{itemize}

%There was a further filter for the removal of repeated jobs, there were many jobs generating several builds that were not several different jobs but several compilations of the same pull request that generated that compilation, because the \travist compiles and generates several tests for that request , in this way it was necessary to remove the repeated data so as not to have repeated information thus generating a false information of the data. Applying the filter in the dataset yielded 1,274 projects, 536,047 Build jobs, 1,295,741 commits, it can be observed that many jobs present in the dataset were the same compliment.

\begin{figure}[ht]
\centering
\includegraphics[width=0.5\textwidth]{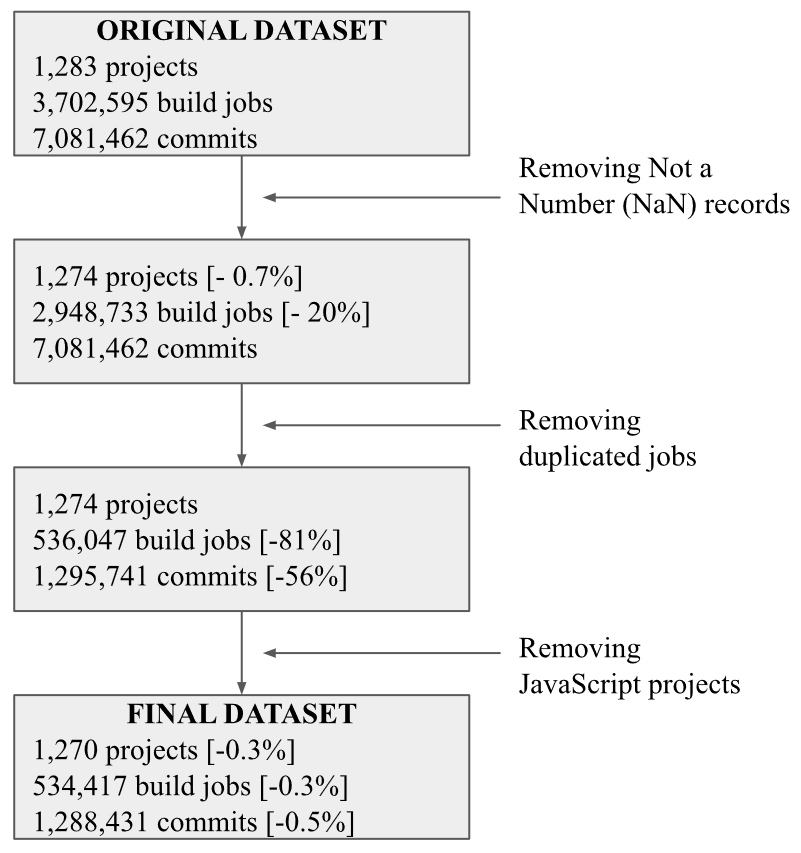}
\caption{The impact of applying each filter on the quantitative of data, and the percentage of reduction.}
\label{fig:filters}
\end{figure}

Figure \ref{fig:filters} present the quantitative of data left after each filter, as well as the percentage of reduction. In the end, we ended up with 1,270 projects, 534,417 build jobs, and 1,288,431 commits. A reduction of 1\% on projects, 85\% on build jobs, and 81\% on commits compared to the original dataset.

% Applying these filters in the dataset, we ended up with 1,270 projects, 534,417 build jobs, and 1,288,431 commits. A reduction of 1\% on projects, 85\% on build jobs, and 81\% on commits compared to the original dataset.

We used this data to provide answers to \textbf{RQ1}, \textbf{RQ2}, \textbf{RQ3}, and \textbf{RQ4}. In particular, for \textbf{RQ2}, since \travist does not provide coverage information, we have to complement it with data from \coveralls. The \coveralls platform tracks the coverage information of software repositories under development on GitHub, GitLab, and BitBucket coding websites. \coveralls has a fine-grained coverage report, comprising each source code file, and each source code line in the file. \coveralls also provides an API in which it made available information about the branch, the total coverage, and the change in the coverage in a particular build. The data available on \coveralls have also be used in other studies (e.g.,~\cite{Hilton:2018:ASE}). Since \coveralls provides an integration with TravisCI, we investigate which projects in the \travist dataset were also configured to use the \coveralls platform. After finding the intersection between these two datasets, we investigate the coverage status on \coveralls of the last available builds of open source available on \travist.

\subsection{Analysing data}\label{sec:analysis}

To help our analysis, we grouped the projects according to their programming language (Ruby and Java) and to their size. In terms of size, we grouped the projects in very small, small, medium, large, and very large. More precisely:

\begin{itemize}
    \item \textbf{Very small}: (less than 1,000 lines of code),  336 projects found;
    \item \textbf{Small}: (more than 1,000 and less than 10,000 lines of code),  622 projects found;
    \item \textbf{Medium}: (more than 10,000 and less than 100,000 lines of code), 261 projects found;
    \item \textbf{Large}: (more than 100,000 and less than 1,000,000 lines of code), 36 projects found.
    \item \textbf{Very large}: (more than 1,000,000 lines of code), only one project found.
\end{itemize}

One may argue that our sample of small projects should be removed from this study. However, although our set of very small projects might not be mission-critical, they are already configured to use \travisci, which makes them valuable for this research. Still, these projects differ from other vary small projects on GitHub that do not use \travisci, which may encompass books and classroom projects. 
Moreover, since we found only one very large project (the \texttt{aws/aws-sdk-java} Java project), when we present the distributions grouped according to the project size, we do not show data for this very large group. We used \travist information to measure lines of code. According to the \travist dataset website, the column ``gh\_sloc'' refers to the ``\emph{Number of executable production source lines of code, in the entire repository}'' %\bnote{continuo achando essa frase avulsa... coluna de quê e onde?}.\gnote{essa coluna e do dataset que usamos. faz mais sentido agora?}

\subsection{Replication package}\label{sec:replication}
For replication purposes, all scripts and code used to deal with the \travist dataset are available as a Jupyter notebook\footnote{https://github.com/wagnernegrao/ci-analysis}.

%in the GitHub repository\footnote{https://github.com/wagnerfns/ci-analysis}.

\section{Results}\label{sec:results}

In this section we report the results grouped by each research question.

\subsection{RQ1: How common is running CI in the master branch but with infrequent commits?}

\begin{figure}[ht]
\centering
\includegraphics[scale=0.38, clip=true, trim= 30px 60px 60px 60px]{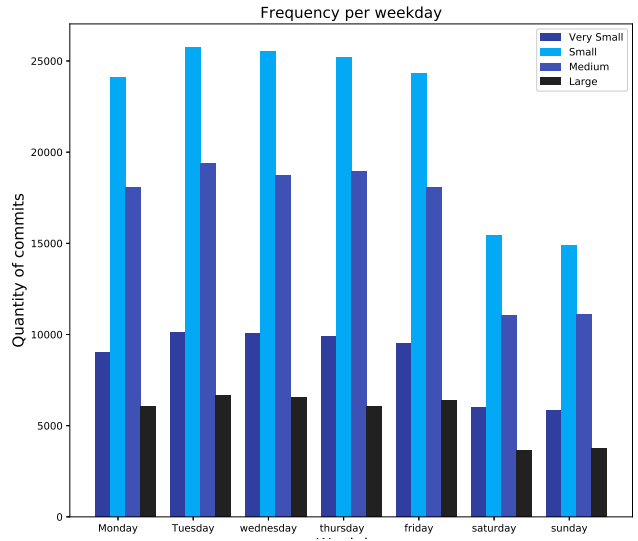}
\caption{Size of the project and its frequency perday of the week.}
%\gnote{essa terceira barra ta quase invisivel. tem como mudar?} %\gnote{essa figura n˜åo ajuda muito. não poderiamos ter a media ou mediana de commits por dia da semana? outra coisa: tem como colocar um espaço entre as 4 barras? colado assim fica dificil entender onde começa e onde termina um grupo}
\label{fig:frequecyPerWeek}
\end{figure}

In this first research question we are intended to analyze infrequent commits made at the master branch. We start by filtering out the commits made to other branches, resulting in a total of 42,3045 commits in the master branch. These commits lead to 368,886 builds in 1,270 open source projects. Figure~\ref{fig:frequecyPerWeek} shows the absolute number of commits according to the week day. Our next logical step was to empirically categorize what are infrequent commits. For each group of projects, we studied the frequency of commits per day. We found out a remarkable uniformity, as Table~\ref{tab:commits} shows.

\begin{table}[!hbt]
\centering
\caption{Information about our dataset (per language)}
\label{tab:commits}
\begin{tabular}{lrrrr}
\toprule
 & average & median & 3rd quartile & standard dev. \\
\midrule 
Very small & 2.13 & 2.0 & 2.0 & 2.15 \\
Small & 2.32 & 2.0 & 2.0 & 2.61 \\
Medium & 2.35 & 2.0 & 2.0 & 2.78 \\
Large & 2.91 & 2.0 & 3.0 & 3.38 \\ 
Very large & 2.68 & 2.0 & 3.0 & 1.62 \\ 
\bottomrule
\end{tabular}
\end{table}

%on average, very small projects have 2.13 commits per day (median: 2.0, 3rd quartile: 2.0, standard deviation: 2.15), small projects have 2.32 commits per day (median: 2.0, 3rd quartile: 2.0, standard deviation: 2.61), medium projects have 2.35 commits per day (median: 2.0, 3rd quartile: 2.0, standard deviation: 2.78), large projects have 2.91 commits per day (median: 2.0, 3rd quartile: 3.0, standard deviation: 3.38), and very large projects have 2.68 commits per day (median: 2.0, 3rd quartile: 3.0, standart deviation: 1.62) %\gnote{complementar}. 

\begin{figure*}[h]
\centering
$
\begin{array}{cc}
     \includegraphics[scale=0.4, clip=true, trim= 0px 0px 0px 0px]{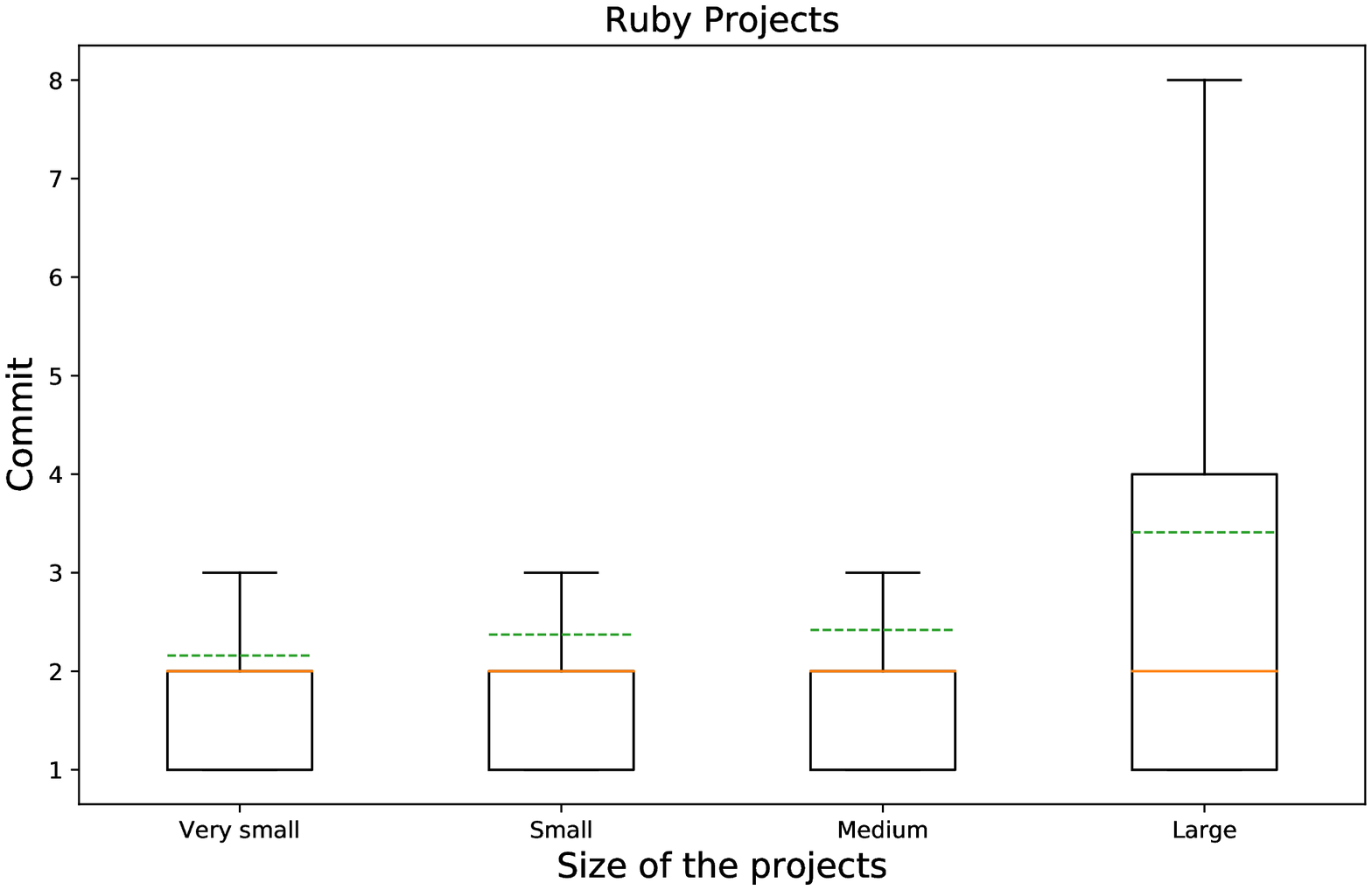} &  \includegraphics[scale=0.4, clip=true, trim= 0px 0px 0px 0px]{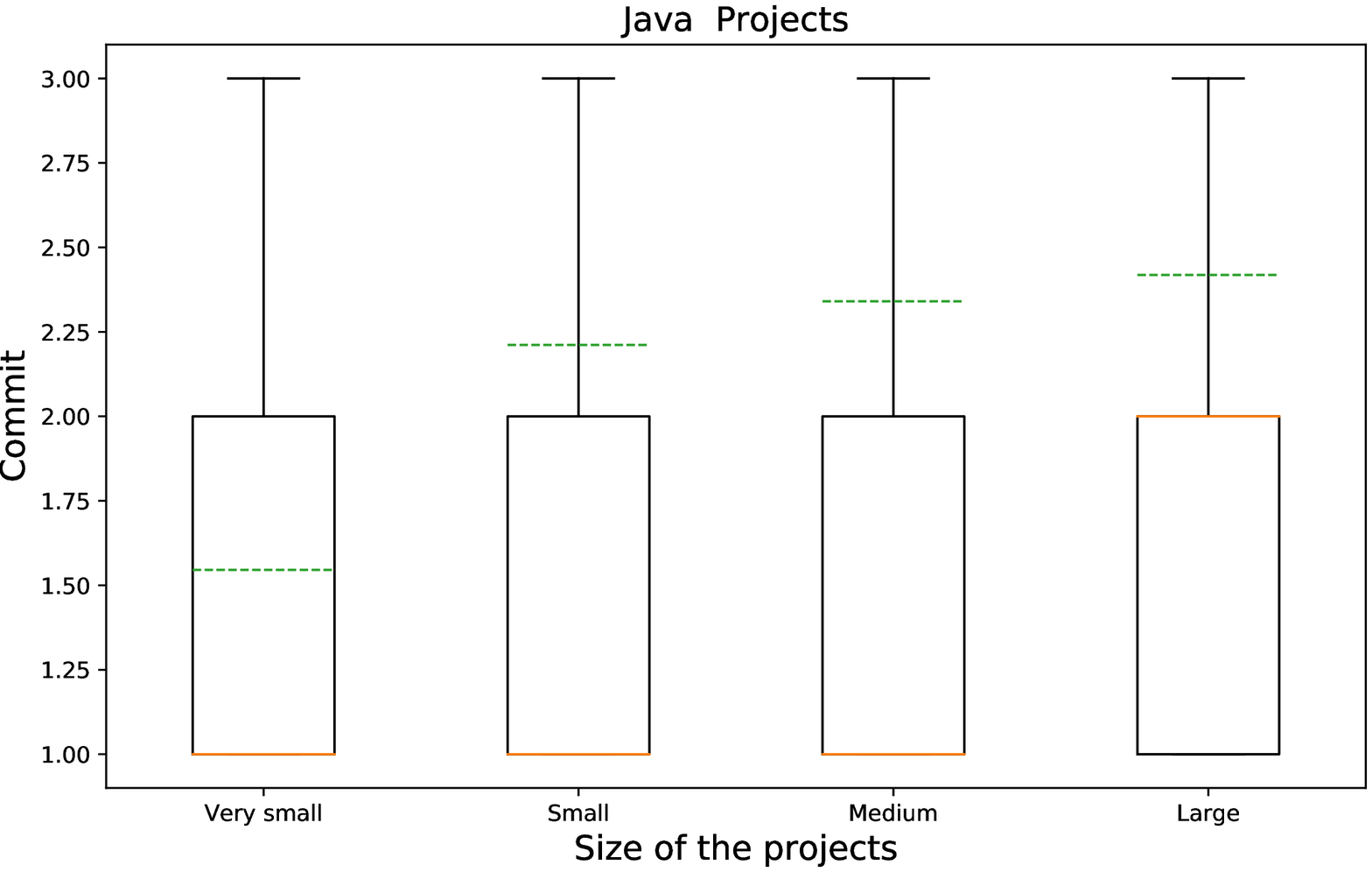} \\
\end{array}
$
\caption{Frequency of commits, grouped by the size of the projects (boxplots), and the programming languages (Ruby on the left and Java on the right). We removed the outliers to ease visualization.}
%\gnote{em cima deve ser ``Ruby Projects'' e em baixo deve ser ``Size of the projects''}}
\label{fig:project_commit_frequency}
\end{figure*}

As one could see, the average number of commits per day is between 2.13 (for very small projects) and 2.91 (for large projects). We also noted a very similar commit frequency when considering the programming language used. For instance, for very small, small, medium, large, and very large Ruby projects, the average of commits per day are, respectively, 2.15, 2.37, 2.42, 3.41, and 0.0 commits. In the Java projects, we found an akin finding: the average of commits per day for the very small, small, medium, large, and very large projects are, respectively, 1.68, 2.15, 2.25, 2.35 and 2.68 commits. Figure~\ref{fig:project_commit_frequency} shows the two distributions. 

Overall, the average of commits per weekday per day is 2.36 (regardless of the size of the project, programming language, and weekday). We then considered a project with infrequent commits any project with an average lower than 2.36 commits per day. %This does not mean, however, that the studied projects have to perform 2.36 commits every single day. In contrast, the project could commit less in the beginning of the week, but commit more in the end of the week.
This empirical observed threshold is somehow in line with the grey literature, which suggest that ``\emph{CI developers must integrate all their work into trunk (also known as mainline or master) on a regular basis (at least daily).}''\footnote{https://continuousdelivery.com/foundations/continuous-integration/} %\bnote{ou seja, poderiamos colocar o threshold como 1 e assim evitariamos usar a media... mas talvez isso tenha alto impacto nos graficos e em todas as intrepretaçoes ... bato nessa tecla pq colocar o threshold na media em uma distribuicao possivelmente normal e com poucos outiliers, vai fazer com que os que ficam a baixo do threshold sejam algo em torno de 50\% sempre, o que tornaria a analise inocua ... algo como dizer que metade estao abaixo da metade. eh uma conclusão obvia}. 

However, when we analyzed how common our studied projects are adhering to this threshold, we found that 748 (59.60\%) face from this infrequent commits concern (214 (56,51\%) Java and 534 (60,89\%) Ruby). Figure~\ref{fig:project_commit_frequency} shows the distribution of commits per day, but now grouping the results in terms of the Ruby and Java programming language. As one can observe, Ruby projects tend to be more active than Java projects (median of commits for Ruby projects is 2.00 and for Java projects it is one). More interestingly, however, is the fact that Java projects have a very stable commit behavior, even when considering projects with different size. In particular, regardless of the size of the Java projects, 50\% of them have infrequent commits. This finding is particularly relevant because if developers take too much time to commit to master (e.g., when working locally or on other branches), they may have to deal with merge conflicts more frequently, which not only require substantial effort from them but also hinder software development activities~\cite{Cavalcanti:2017:OOPSLA}

However, this finding does not ring true when considering Ruby projects. Large Ruby projects, in particular, tend to be more active than the other ones.

%On the other hand, only 507 of the studied projects (164 Java projects and 343 Ruby projects) have more than 2.36 commits a day. %It is important to note that these projects should have this commit activity on every weekday.

%In order to analyze the infrequency in thelanguages, the projects were divided into three groups, being very small, small, medium and large. The frequency in the Ruby language for small projects is defined as, very small projects have a mean of  2,15 commits, small projects must have average of 2.37 commits, average projects must have commits of 2.42 commits and large projects have commits in average 3.41, Figure~\ref{fig:ruby_project_commit_frequency}. %The frequency in the Java language, very small projects have a mean of  1,68 commits, small projects have a mean of 2.15 commits, average projects have an average of 2.25 commits and large projects have an average of 2.35 commits. All projects that are below the average of each type of project are considered infrequent, Figure~\ref{fig:java_project_commit_frequency}.

\MyBox{\textbf{RQ1 Summary:} We categorized projects with infrequent commits when they have less than 2.36 commits per day. We found that, in general, $\sim$60\% of the projects in our dataset suffer from infrequent commits. In particular, half of the Java (regardless of their size) have infrequent commits, which may hinder software development activities.}\\

\subsection{RQ2: How common is running a build with poor test coverage?}

In this research question, we are interested to understand the test coverage of our studied projects. If the test coverage is small, it may suggest that the use of \travisci is underused, since the potential benefits of running a comprehensive test suite automatically to find bugs is skipped.

\begin{figure}[ht]
\centering
\includegraphics[scale=0.6, clip=true, trim= 0px 0px 0px 0px]{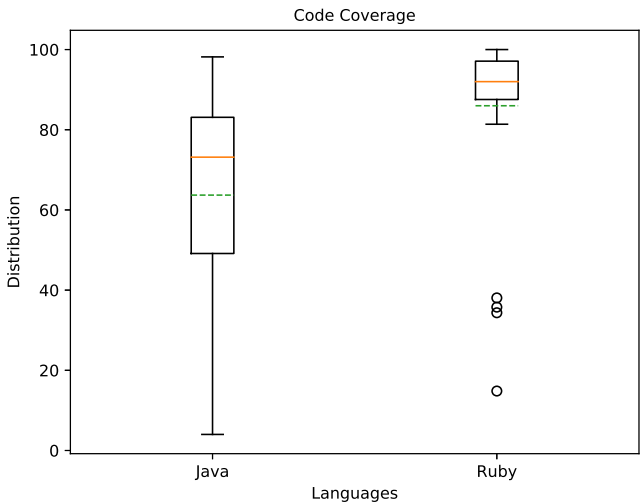}
\caption{Code coverage per programming language.}
\label{fig:codeCoverage}
\end{figure}

From our corpus of 378 Java projects and 877 Ruby projects, we found only 25 Java projects and 58 Ruby projects with \coveralls information. This reduced our corpus to 83 projects. There is a gotcha, however. The last release of the \travist dataset was on 2017. We then applied another filter to select only projects with coverage information during the same period that we had build information. More concretely, we selected the last build record available on \travist and tried to match whether \coveralls had coverage information on the same day of the last build.
%,  whereas \coveralls can be integrated with \travist. 
Since we observed that the relationship between build records on \travist and coverage records on \coveralls is roughly one to one, we provide a grace period: for those projects that we did not find coverage information for the exact same day of the last build, we extended our search to find coverage records over the last seven days prior to the build day. For instance, if the last build information that we have for a given project is on November, 20th 2016, we first search for coverage information on the same day (November, 20th 2016); if no data was found, we search for coverage information until November, 13th 2016. After this process, we ended up with 16 Java projects and 35 Ruby projects. Overall, the average coverage of these projects in the last available build was 78.99\% (median: 88.46\%). Figure~\ref{fig:codeCoverage} shows the coverage distribution for these two set of projects. 

As we can see, the coverage of these group of projects varied greatly. On one hand, Java projects seem to have much more coverage variation. On average, Java projects have 63.69\% of code coverage (median: 73.16\%) 3rd quartile: 83.10\%, standard deviation: 27.01\%), varying from 4.0\% at the lowest coverage, up to 98.17\% at the highest coverage. The Java project with the lowest coverage is \texttt{connectbot/connectbot}. Moreover, we found three additional Java projects with less than 50\% of coverage rate, namely: \texttt{psi-probe/psi-probe} (24\% of coverage), \texttt{myui/hivemall} (33\% of coverage), and \texttt{igniterealtime/Smack} (35\% of coverage). For these projects, we conducted a follow up analysis to understand whether their coverage evolved over time. Interestingly, we observed that these projects did not expressed major changes in their level of coverage. For instance, \texttt{psi-probe/psi-probe} improved from 24\% in 2016 to 35\% of coverage in 2019, \texttt{myui/hivemall} kept the same coverage level in 2019 as from 2016: 33\%, and \texttt{igniterealtime/Smack} improved from 35\% in 2016 to 38\% coverage in 2019.  Still, regarding \texttt{connectbot/connectbot} which is the Java project with the lowest code coverage (4\%), we observed that this project improved its coverage to 34\%. In particular, we identified one single commit\footnote{https://github.com/connectbot/connectbot/pull/410/commits/575766a6444} that made the coverage jumped from 4\% to 29\%. When we inspected this particular commit, the commit message suggested that the intention was to ``\emph{Create combined coverage target}''. Inspecting the commit changes, we observed that the author of this commit decided to exclude some directories (that may contain code not relevant to this project) from the build process. After applying this commit, the coverage improved 25\%.
On the Ruby side, however, the landscape is completely different. We observed that, on average, the coverage of the Ruby project is 85.98\% (median: 92\%, 3rd quartile: 97.10\%, standard deviation: 20.93\%), varying from 14.83\% at the lowest coverage, up to 100\% at the highest coverage. More interestingly, however, is the fact that 17 (48\%) Ruby projects have coverage greater than 90\%. 

We hypothesize that this high coverage scenario for Ruby projects is intrinsically related to the characteristics of the Ruby programming language. Since Ruby is a dynamic typed programming language, developers are only aware of eventual bugs caught by the type system during runtime. Therefore, they may have to rely on a good test suite to minimize eventual bugs that may only appear on the fly. On the other hand, Java developers take advantage of static typing, which avoid some class of bugs that could pass through unattended otherwise.% \bnote{it's a good point, but... alguma referencia pra embasar isso? se tiver, melhor}.

%To perform the coverage measurement we took each build of the travis dataset and searched for the corresponding coverage information on the coveralls being recorded no later than 7 days after the build date found on Travis. After that we selected 3 coverage information for each project, with two ends of dates and one being half of them, and averaged 3. For projects that did not contain 3 coverage information or more, we only used the available ones. The reason for using 7 days as a metric limit was the fact that we found, by selecting 10 random projects, that their coverage did not change as much over the 7-day period, The maximum variation recorded was 1.14\%.

%\merge{We will classify that a project has poor coverage when that rate is below 85\%, in Marick's\lnote{discutir} work it is quoted that some respected companies use that number as the basis for good coverage.}

\MyBox{\textbf{RQ2 Summary:} We found 51 projects in our dataset that have records on \coveralls. Although the overall coverage was 78\%, the coverage of Java and Ruby projects differs greatly. The average code coverage of Ruby projects was 86\%, whilst for Java projects it was 63\%. This suggests that although poor test coverage exist, a significant number of studied projects take care of their code coverage.}

\subsection{RQ3: How common is allowing the build to stay broken for long periods?}

To analyze this research question, we studied the period, in terms of days elapsed, of broken builds. For each broken build, we counted the number of days between the commit that broken the build until the commit that fixed the build. Since practitioners did not have a clear rule of thumb for the maximum duration that a build could stay broken (the grey literature suggest that the build should be fixed right away~\cite{fowler:CI}, we took a conservative approach and used the third quartile of the overall duration of broken builds. Therefore, we assume four days as the threshold for this research question (mean: 7 days, 3rd quartile: 4 days, standard deviation: 29 days). 

\begin{figure*}[h]
\centering
$
\begin{array}{cc}
      \includegraphics[scale=0.4, clip=true, trim= 0px 0px 0px 0px]{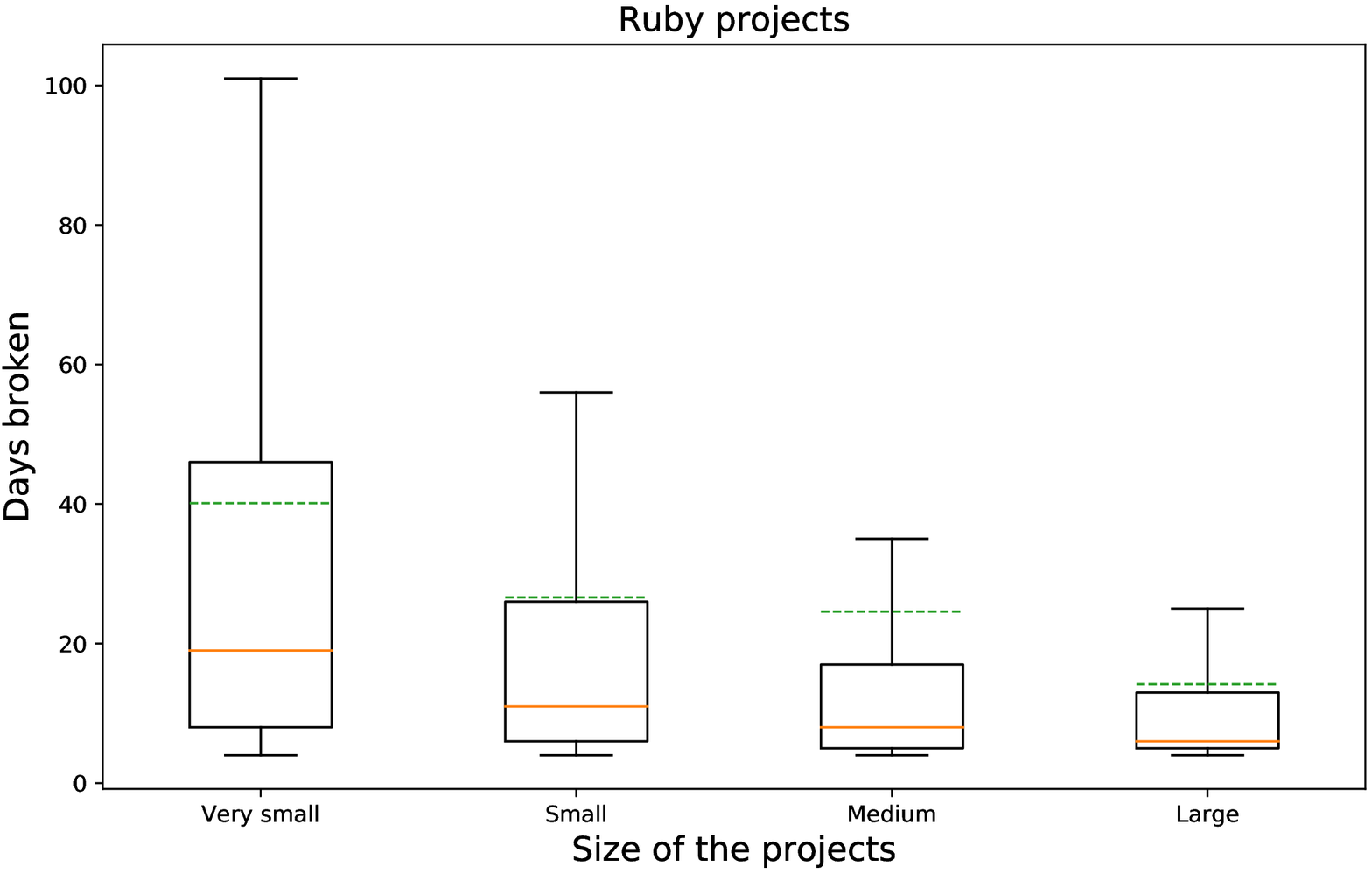}
      &
      \includegraphics[scale=0.4, clip=true, trim= 0px 0px 0px 0px]{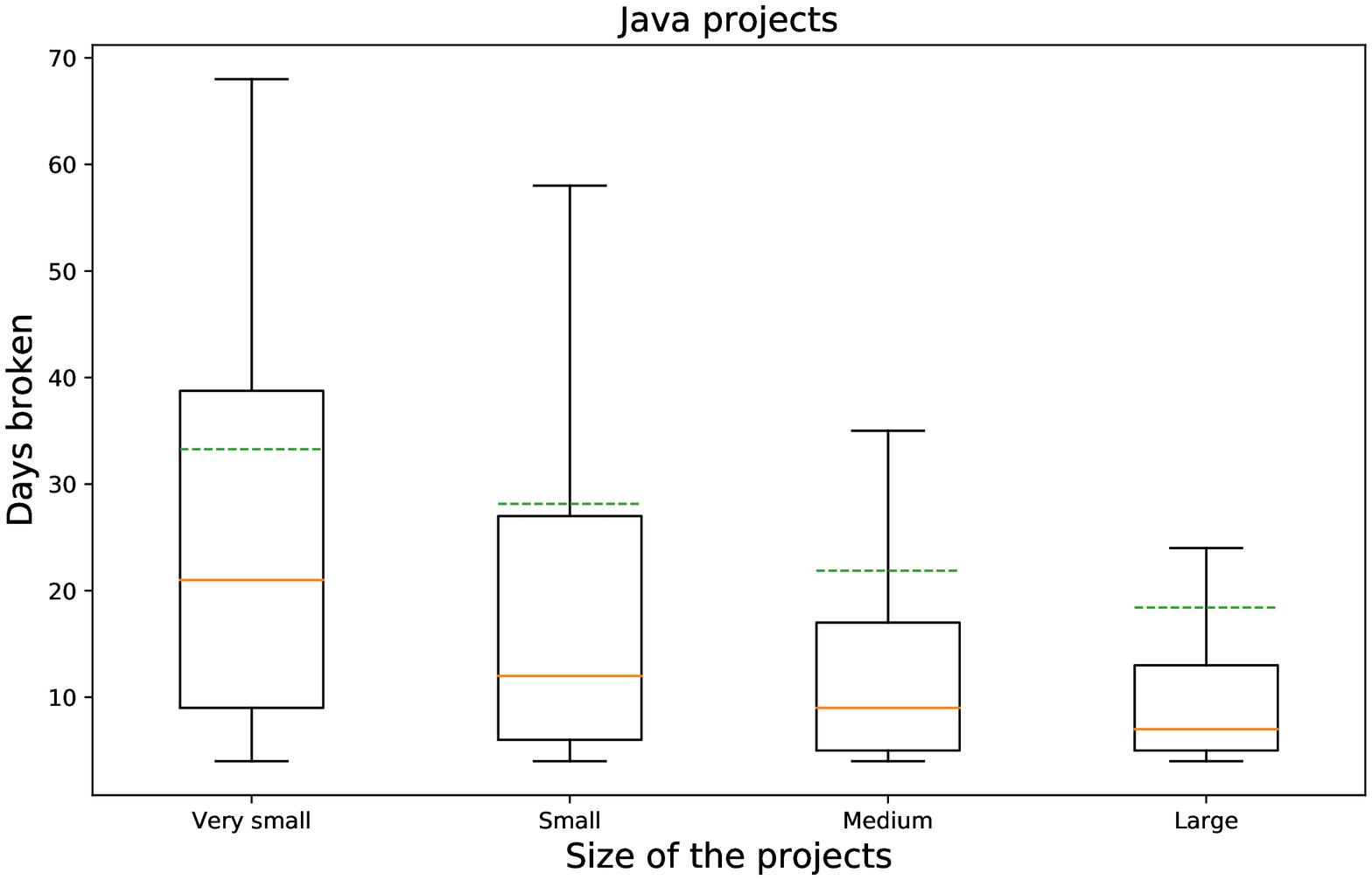}
\end{array}
$
\caption{Distribution of days with broken build, grouped by the size of the projects (boxplots), and the programming languages (Ruby on the left and Java on the right). We removed the outliers to ease visualization.}
\label{fig:days_broken}
\end{figure*}

When we applied this threshold in the dataset, we observed that 1,072 (85.4\%) %\bnote{novamente, o problema de usar como threshold uma medida derivada da amostra ... o terceiro quartil abarca 75\% e o percentual q obtivemos eh algo relatiamente proximo, 85\% - a menos que a distribuiçao seja muito distante da normal, muito skewed, ou com outliers muito pesados, mas nesses casos, de toda forma, as medidas centrais (media, mediana e quartis) perdem o significado} 
out of the the 1,270 projects have at least one broken build that took more than four days to be fixed. Figure~\ref{fig:days_broken} shows the distribution. More concretely, 85.42\% of Java projects have at least one long-to-be-fixed broken build (88,48\% for Ruby projects). However, the most interesting observation for this set of experiment is related to broken builds according to the size of the projects. In contrast to a natural belief, large projects (which tend to be more complex and difficult to reason about) are the ones that fix a broken build faster. This holds true for large projects written in the two programming languages, and when comparing to every other size of projects. More specifically, large Java projects let build stay broken, on average, for 2 days (median: 0 days, 3rd quartile: 1 days, max: 408 days). For large Ruby project, the average is 1 day (median: 0 days, 3rd quartile: 1 days, max: 140 days). This finding is in sharp contrast to what was found in smaller projects. For instance very small Java projects, on average, let the build to stay broken for 20 days (the average for very small Ruby projects: 21). 

One hypothesis for this behavior is that large projects may count with a large workforce of source code contributors that are readily available to fix broken changes. Moreover, large projects may have a very large user base; for this reason, a broken build may impact several users. Conversely, small projects not only may have to rely on a single code contributor~\cite{Avelino:2016:ICPC} but also may not be as popular as large projects (therefore, there might be little rush to fix a broken build since a very small user base would be regularly updating the master).

\MyBox{\textbf{RQ3 Summary:} We observed that 85\% of the analyzed projects have at least one build that took more than four days to be fixed. Interestingly, we observed that large projects (either Java or Ruby) have less long broken builds than smaller projects.}

\subsection{RQ4: How common are long running builds?}

%To know how long a build might have been, the duration time was analyzed, the value of 368,886 builds builds of the dataset was removed with values NaN (Not a Number), thus presenting 55,044 builds for analysis.

\begin{figure*}[h]
\centering
$
\begin{array}{cc}
      \includegraphics[scale=0.4, clip=true, trim= 0px 0px 0px 0px]{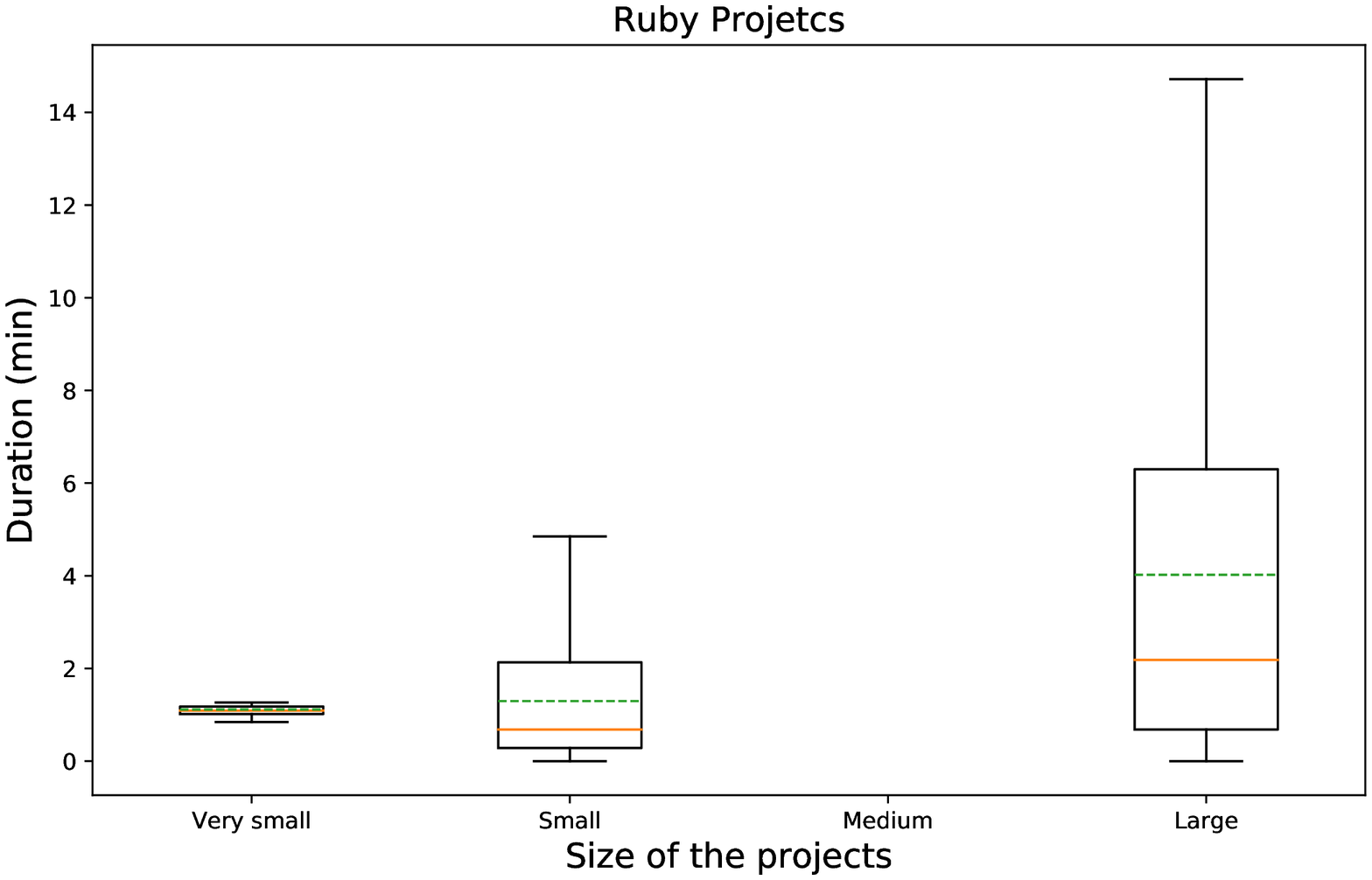}
      & 
      \includegraphics[scale=0.4, clip=true, trim= 0px 0px 0px 0px]{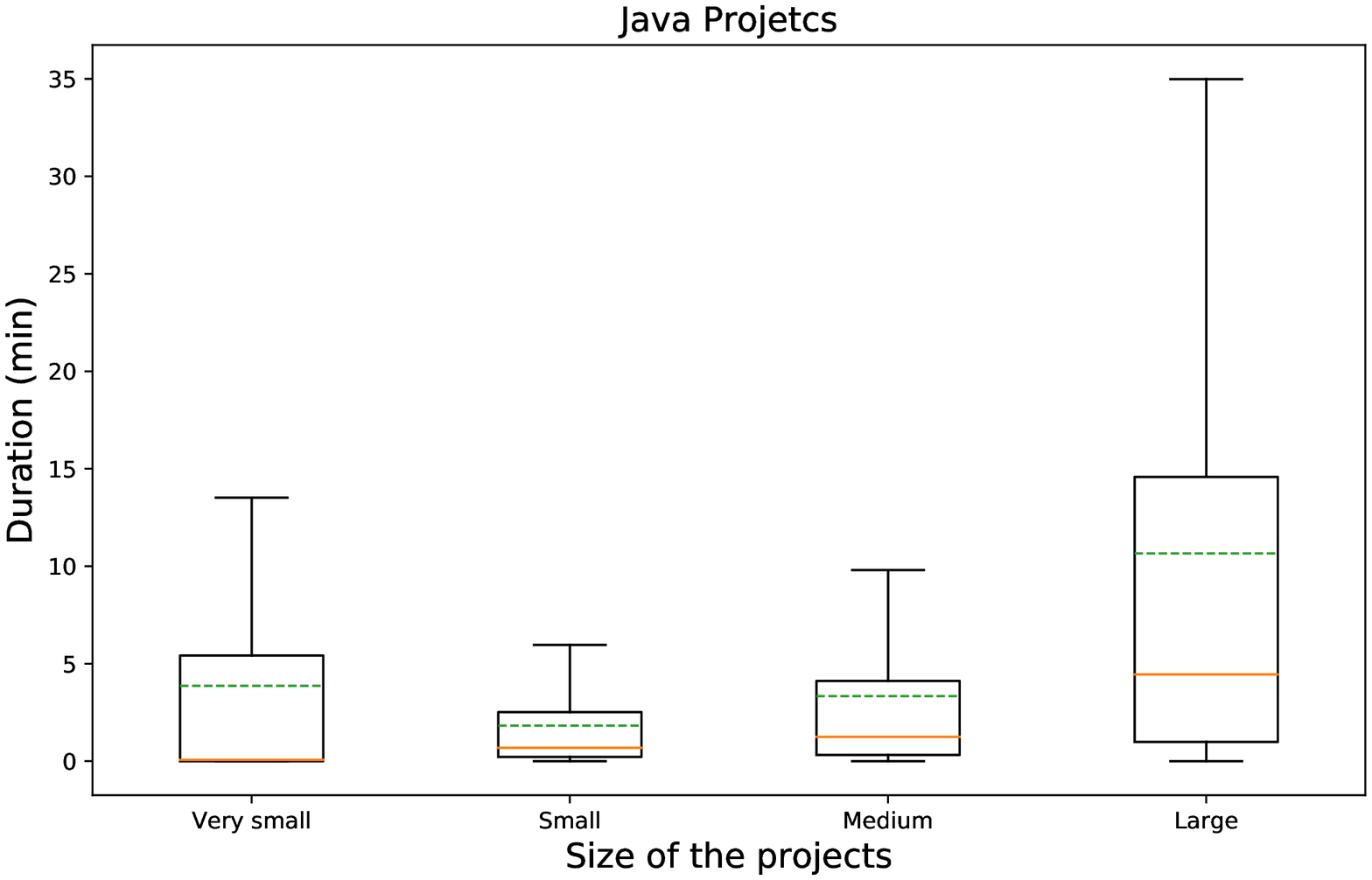} \\
\end{array}
$
\caption{Build duration, grouped by the size of the projects (boxplots), and the programming languages (Ruby on the left and Java on the right). We removed the outliers to ease visualization.}
\label{fig:build_duration}
\end{figure*}

For this research question, we studied the duration of the 368,886 builds in our dataset. To measure the build duration, we relied on the ``tr\_duration'' column of the \travist dataset that ``\emph{The full build duration as returned from the Travis CI API}''. When investigating the data, we perceived several NaN records in this particular column. We then removed all NaN records, which reduced our data set from 368,886 builds to 55,044 builds. This new sample comprehends 261 projects (253 Java and 8 Ruby). Among the Java projects, there are 12 very small, 98 small, 118 medium, and 24 large ones. 
%\merge{a soma da 252 pq 1 projeto e very large}. 
For Ruby projects, we found 1 very small, 6 small, and 1 large. No medium and very large Ruby projects were found in this regard.

Figure~\ref{fig:build_duration} shows the distribution of the builds' time duration.
For this new set of projects, we observed that, on average, the build took 4 minutes and 18 seconds to run (median: 1 minute and 26 seconds, 3rd quatile: 4 median and 42 seconds, standard deviation: 7 minutes and 26 seconds). Moreover, the build of projects written in Java take on average 4 minutes and 25 seconds (median: 1 minutes and 24 seconds, 3rd quartile: 4 minutes and 31 seconds, standard deviation: 7 minutes and 49 seconds), whereas the the build of projects written in Ruby take on average 3 minutes and 40 seconds (Median: 1 Minute and 47 seconds, 3rd quartile: 5 Minute and 43 seconds, standard deviation: 4 minutes and 23 seconds). 
More interesting to this research, however, are the time duration of builds made in the large projects, either for Ruby or Java projects. We found 43 (16\%) out of the 261 projects have at least one build that took longer than 10 minutes. These projects have produced 7,046 long builds out of the 55,044 total ones.

One natural thought is that large projects may take more time to build because they have a more complicated compilation process. To shed some light along these lines, we investigated the build process of some large projects. We found two large Ruby projects with build time longer than 10 minutes. One of these projects is \texttt{jruby/jruby}, which is an implementation of the Ruby programming language for the Java Virtual Machine. When we analyzed the build output for this project, we noted it took about 18 minutes to run integration testing suite (which has 3,389 tests). %\lnote{A m\édia da dura\ç\~ao da build desse projeto \'e 803 segundos (cerca de 13 minutos)}

In contrast, we found 43 Java projects with long builds. At the worst case scenario, we found one project, \texttt{geoserver/geoserver}, which its build took 59 minutes. When analyzing its \travisci configuration file, we observed that it downloads the maven binaries, and execute its for every build. This process requires the CI system to download third-party libraries used in the project during every single build. Other projects with long build duration are \texttt{facebook/presto}, \texttt{spotify/helios}, and \texttt{biojava/biojava}.

\MyBox{\textbf{RQ4 Summary:} We observed that only 16\% of the projects do not adhere to the 10 minutes rule of thumb for build duration. However, when considering Java large projects, the landscape changes significantly: 52\% of them take more than 10 minutes.} %\bnote{talvez isso seja um esparado, ja que projetos maiores tendem a demorar mais para compilar}}

\section{Threats to Validity}\label{sec:threats}

In a study such as this, there are always several limitations and threats to validity. First, this research was built upon the \travist dataset. Although this dataset provide a comprehensive taxonomy about build information of over one thousand of GitHub projects that use \travisci, the last release of this dataset was on 2017, and the most recent build was recorded on March, 21st, 2016.  Therefore, our results cannot be extended to the build behavior of these projects today. However, due to the scale of our analysis, we do not expect major changes in the main results, if the most recent builds were considered.

Moreover, to answer RQ1 and RQ3 we extract thresholds from our sample and apply them to the sample itself. This approach may produced self-evident conclusions in case of normal or close to normal distributions. However, we still decided to proceed with this strategy since there is no golden standard, in the context of CI, about what is an adequate number for the frequency of commits nor an acceptable period of time for builds to remain broken. We expect that these questions could be revisited in future works.

Further, we used \coveralls to gather coverage information. Since this is a proprietary third-party service, we have to blindly rely on its output. A possible mitigation plan would be to download, compile, and execute tests for the open-source projects locally. However, this is often non-trivial task (e.g., some projects fail, some projects require manual configuration, etc). Since recent related work is also employing \coveralls (e.g.,~\cite{Hilton:2018:ASE}), we opted to use the \coveralls infrastructure to gather coverage information for this work as well. 

Another threat to validity is related to the amount of NaN (Not a Number) records in our dataset. To avoid influence the results with these NaN records, we decided to removed them all. This decision, however, may also affect some of our findings. For instance, since there were several NaN in the build duration column, we ended up without medium Ruby projects for RQ4. Finally, one may argue that our approach of providing a grace period could introduce bias, since seven days of software development can greatly change the coverage information. To mitigate this concern, we investigate a random sample of 10 projects and we perceived that their coverage do not change much during the period of seven days. The maximum variation recorded was -1.14\% (two projects also had zero variation).

%Another problem encountered was when relating the dataset of travis with coveralls, because coveralls does not have accurate information about the build to which it is related, so it can not accurately map data from travis to coveralls, since you only have the name of the project and the day of the build to make the relation.}

\section{Related Work}\label{sec:relatedwork}

There is a recent flow of empirical studies targeting continuous integration systems, in general, and TravisCI, in particular. 

Vasilescu and colleagues~\cite{Vasilescu:2014:CIS} performed a quantitative study of over 200 active Github projects. They restricted their search to Java, Ruby, and Python projects. Among the findings, they found that 92\% of the selected projects have configured to use Travis-CI, but 45\% of them have no associated builds recorded in the Travis database. 
%They also found that direct contributions (pushed commits) are more frequent than pull-requests. 
Differently from Vasilescu and colleagues~\cite{Vasilescu:2014:CIS}, our work focuses on analyzing projects that, despite using CI tools,
do not actually employ CI practices. We also search for builds involved in several other programming languages. 

The study by Hilton and colleagues~\cite{Hilton:2016} aimed at understanding how software developers use CI tools. 
%More specifically, why do some projects do not use CI or how CI configuration evolve. 
Through the analysis of CI builds and a survey, they observed that CI is widely adopted in popular projects and reduces the time between releases. Our work complements their work by analyzing broken builds. Although the work of Hilton and colleagues~\cite{Hilton:2016} provide some initial discussion about build breakage, they did not provide an in-depth investigation in this regard.

The work of Vasilescu and colleagues~\cite{Vasilescu:2015:QPO} analyzed historical data of GitHub projects to see the effects of using CI. They observed that CI helped to increase the number of accepted pull requests from core developers. They also found that CI reduces the quantity of rejected pull-requests while maintaining code quality. Our work complements the previous research by showing that not all projects that adopt CI tools actually employ the CI practices. 

Beller and colleagues~\cite{Beller:2017:Oops} studied ``how central testing really is in Continuous Integration''. They observed that testing is the main reason as to why builds fail in CI.
%while creating tests takes a considerable part of the developer's time, they rarely execute them in their IDE and might instead offload them to the Continuous Integration platform, which is in line which some of the findings of our study. 
Different to their work, we also focus on the frequency of commits, build duration, and the required time to fix a broken build. 
%which is artifact-focused (\emph{i.e.,} based on data acquired from software repositories), our work is developer-focused (\emph{i.e.,} based on the perception of software developers that have experience with CI tools). 

Taher Ghaleb and colleagues~\cite{Ghaleb:EMSE:2019} studied the CI builds with long durations. They built a mixed-effects regression model to study 67 GitHub projects with long build durations. Among the observations, the authors highlight that some CI practices can produce a longer build duration. 
%For example, configuring the CI service to check, in every build run, for dependency updates can generate an unnecessary overhead in the building process. 
Complementary to the work by Taher Ghaleb and colleagues~\cite{Ghaleb:EMSE:2019}, we also study the usage of CI in builds with a long duration. However, we also study other CI usage scenarios, such as the use of CI with infrequent commits and poor testing coverage.    

%Other studies have focused on empirically checking the potential benefits of CI in software projects. 
Bernardo and Colleagues~\cite{Bernardo:MSR:2018} empirically studied whether the adoption of Travis-CI is associated with a shorter time to deliver new functionalities to end users (i.e., delivery delay). They found that the adoption of Travis-CI may not always quicken the delivery of software functionalities. However, they observed that adopting CI is usually associated with a higher proportion of functionalities delivered per software release. We complement their work by quantitatively studying unhealthy CI practices. For example, the observation that adopting CI may increase the time to deliver functionalities might be associated with some of the CI bad practices that we have studied in our work (e.g., poor code coverage). 

Zhao and colleagues~\cite{Zhao:ASE:2017} empirically investigated the adoption of Travis-CI in a large sample of GitHub projects. They quantitatively compared the CI transition in these projects using metrics such as commit frequency, code churn, pull request closing, and issue closing. In addition, they conducted a survey with a sample of the developers of the studied projects. The survey consisted of three questions related to the adoption of Travis-CI and CI in general. The main observations were: (i) a small increase in the number of merged commits after CI adoption; (ii) a statistically significant decreasing in the number of merge commit churn; (iii) a moderate increase in the number of issues closed after CI adoption; and (iv) a stationary behavior in the number of closed pull requests as well as a longer time to close PRs after the CI Adoption. Contrary to the work performed by Zhao and colleagues~\cite{Zhao:ASE:2017}, we studied four scenarios of unhealthy CI usage instead of the impact that the adoption of CI may bring to a software project.

The study by Maartensson and colleagues~\cite{Maartensson:JSEP:2019} investigated the following general question ``{\em How can the continuous integration and delivery pipeline be designed in order to support all existing stakeholder interests}.'' To this end, the authors surveyed practitioners from 10 software development companies which develop large‐scale software‐intensive embedded system.They proposed a conceptual model that shows practitioners how better design a CI pipeline to include test activities that support all the different interests of the involved stakeholders. Differently from their work, we qualitatively study 1,270 open-source projects. Our work can be complementary to the work by Maartensson and colleagues~\cite{Maartensson:JSEP:2019} in the sense that we shed light on some bad CI practices that can be avoided when designing a CI pipeline. 

Gallaba and colleagues~\cite{Gallaba:ASE:2018} empirically investigated the noise and heterogeneity that might lurk in CI build data. They found that CI builds may contain breakages that are ignored by developers. Their analyses of Java projects reveal that builds may contain breakages that occur outside the build tool. Instead of studying the possible noise in CI build data, our work focuses on the unhealthy usage of CI. Our work can complement the study of bias in the existing quantitative analyses, since we observe that not all usage of CI can be healthy. For example, many CI projects that have been quantitatively studied in the field may contain unhealthy CI usage scenarios. 

Finally, Zampetti and colleagues~\cite{Zampetti:SANER:2019study}
studied the interplay between pull request reviews and CI builds. They analyzed a sample of 857 pull requests that incurred in a build breakage when they were submitted. The result of this analysis was a taxonomy of build breakage types that are discussed through pull requests. They also surveyed 13 developers to complement the observations of their qualitative study. 11 out of the 13 respondents highlighted that the build status actually contribute to the decision taken by the pull request reviewer. Also, the respondents mentioned that the majority of reviewers do not accept a pull request if the build is failing. Our study complements the work by Zampetti and colleagues~\cite{Zampetti:SANER:2019study}, since we observe that around 60\% of our studied projects perform infrequent commits, which makes the merging process harder. 
%Other studies focus on understanding the problems that CI users may face when dealing with broken. For example, builds~\cite{Pinto:CHASE:2017,Pinto:SPE:2018}, how do the developers' involvement influence the build
%status~\cite{Reboucas:2017:CII}, how changes impact the build in different operating systems and runtime environments~\cite{Zolfagharinia2017}, and even the interplay between non functional requirements and the build status~\cite{paixao2017}. 
%Some studies have reported the practitioners perception regarding a broken build~\cite{Pinto:SPE:2018,Pinto:CHASE:2017}
%Other studies have proposed approaches to minimize build breakage~\cite{Ananthanarayanan:2019:EuroSys}
%To the best of our knowledge...

\section{Conclusions}\label{sec:conclusions}

In this work, we studied four bad practices that comprehends our notion of Continuous Integration Theater, namely (1) performing infrequent commits to the mainline repository, (2) building a project with poor test coverage, (3) allowing the build to remain in a broken state for long periods, and (4) using CI with long duration builds. To perform our empirical study, we leveraged the \travist dataset. In addition, whenever necessary, we used \coveralls to gather the test coverage of our studied projects. Through the study of 1,270 projects, our results reveal that although some bad practices are commonly employed, such as infrequent commits in the master branch (in $\sim$60\% of the projects), other bad practices are not as frequent (such as a build taking too long to process). 

Our research shows that the `CI Theater' is present, to some extent, in a considerable amount of software projects. This results imply that existing research that analyzes the benefits of CI (e.g., the time to deliver new functionalities) should consider whether the studied projects have also adopted good CI practices. Using projects that perform the `CI Theater' in empirical analyses may introduce some bias in the analyses. 

For future work, we plan to extend our analysis to a newer dataset of CI builds to verify whether the 2016 data generates a significant impact our the results. Still, we plan to interview developers to investigate the effects of bad practices on software health. Finally, we plan to enrich the list of bad practices either by asking practitioners other bad practices that they face with CI or by empirically observing developers working with CI.

\vspace{0.2cm}
\noindent
\textbf{\textit{Acknowledgments}.} We thank the reviewers for their helpful comments. This research was partially funded by CNPq/Brazil (406308/2016-0) and UFPA/PROPESP.

\bibliographystyle{unsrt}
\bibliography{references}

\end{document}